# Supercooled Liquid Water Diffusivity at Temperatures near the Glass Transition Temperature

R. Scott Smith, Wyatt A. Thornley, Greg A. Kimmel, and Bruce D. Kay

*Physical Sciences Division, Pacific Northwest National Laboratory, Richland, Washington 99352*

**Corresponding Authors**

R. Scott Smith, Email: Scott.Smith@pnnl.gov

Bruce D. Kay, Email: Bruce.Kay@pnnl.gov

**Abstract**

Isotopically layered amorphous solid water films were used to measure the diffusivity of deeply supercooled liquid water near the glass transition. The films, composed of separate $H_2^{18}O$ and $H_2^{16}O$ layers, were grown by vapor deposition at low temperature and then heated to observe the intermixing of the isotopic layers. Very slow heating rates (as low as $10^{-4}$ K/s) were used to decouple the diffusion and crystallization processes to ensure that the observed intermixing occurred at temperatures that were well-separated from the onset of crystallization. Numerical simulations of the desorption spectra were used to extract the translational diffusivities. The diffusivities obtained in this paper are consistent with translational liquid-like motion at temperatures near and above the proposed $T_g$ of 136 K. These findings support the idea that the melt of amorphous water, above its glass transition temperature is thermodynamically continuous with normal supercooled liquid.

**Table of Contents image**

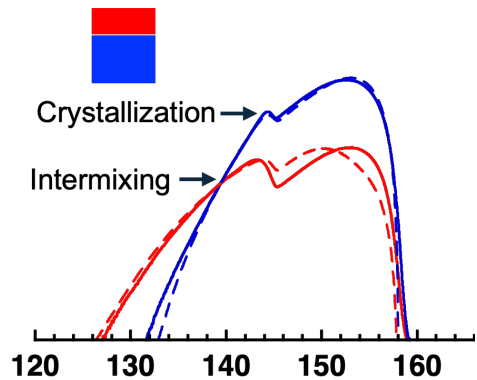

# I. Introduction, Results, and Discussion

Among water's numerous unusual properties is the unresolved question of whether the behavior of amorphous water above its apparent glass transition at ~136 K is that of a true supercooled liquid.[1-4] The answer is important for determining if there is thermodynamic continuity between the melt of amorphous water and normal liquid water.[5-9] The glass transition, which is a kinetic transition not a thermodynamic phase transition, is typically defined to occur when molecules are moving too slowly to sample the available configurational space on a laboratory timescale (~100 s). At the glass transition temperature, $T_g$, the supercooled liquid's structure is effectively "frozen" as an amorphous solid. When an amorphous solid is heated above its $T_g$ the previously inaccessible configurational degrees of freedom become accessible and the rotational and translational properties of the supercooled liquid reemerge.[10-12]

The formation of amorphous water, as for most liquids, requires cooling rates that are fast enough to circumvent crystallization.[1,11,13,14] For water, this can be accomplished by vapor deposition onto a low temperature (<~120 K) substrate.[9,13,15,16] Calorimetric characterizations of amorphous water films report an increase in the heat capacity of ~2 J/K·mol near 136 K which was interpreted as being due to the glass transition.[17-20] This change in heat capacity is relatively small compared to the change at the melting point where the heat capacity increases by ~38 J/K·mol. In fact, the increase in the heat capacity from crystalline ice to liquid water is about a factor of two. Some have argued that the reported glass transition is too weak to be from the unfreezing of translational and orientational (rotational) degrees of freedom expected at the glass transition and that these results may be due to reorientation transitions only.[2,21,22]

Beyond changes in thermodynamic properties, the transformation to a supercooled liquid at the glass transition should also result in the onset of molecular translational motion. Fisher and Devlin conducted infrared experiments to observe isotopic exchange in amorphous water films at ~125 K and concluded that the molecular motion that develops at the glass transition temperature is due to orientational (rotational) diffusion.[23] Similarly, Shepard and Salzman conducted calorimetric and x-ray diffraction experiments and also concluded that their results support molecular reorientation and not translational motion at the proposed $T_g$.[24]

In prior work, we used isotopically layered water films to determine the diffusivity of the melt of amorphous water above its $T_g$ and prior to crystallization.[15,25,26] The time evolution of the intermixing was determined by monitoring desorption kinetics from of the outer layer of the film. We found that when heated above its $T_g$, the intermixing of the isotopic layers was consistent with long-range molecular translation characteristic of liquid-like behavior. However, later worked showed that crystallization, which occurred in concert with the intermixing, strongly influenced the results such that they were not representative of translational diffusion in the liquid.[27,28]

In this letter, we revisit the intermixing of isotopically layered amorphous water films using low temperatures and extremely slow heating rates to decouple the diffusion and crystallization processes. This allows us to measure and analyze the intermixing of the amorphous layers without the interference of crystallization. We find liquid like translational diffusion near the glass transition. The diffusivity has an Arrhenius temperature dependence with an activation energy of ~36 kJ/mol.

Evidence for liquid-like translational diffusion in amorphous water films at or near 136 K is displayed in Figure 1. Shown are the temperature programmed desorption (TPD) spectra for three amorphous water films composed of 20 layers of $H_2^{18}O$ deposited on top of 20 layers of $H_2^{16}O$ at 70 K. Amorphous water formed by vapor deposition on a cold substrate is referred to as amorphous solid water (ASW).[13,16] After deposition, the composite films were heated to and held at 136 K for various wait times (0 s, 3600 s, and 10800 s). After the specific wait time, the films were cooled below 70 K and then heated at a linear rate of 0.5 K/s until desorbed. The bottom set of curves is for the 0 s wait time at 136 K experiment. Initially, desorption is observed only from the top layer ($H_2^{18}O$, blue curve) until about 154 K where the onset of desorption from the bottom layer ($H_2^{16}O$, red curve) becomes apparent. At ~159 K there is a decrease in the desorption rates from both layers that results in an apparent "bump" in the TPD spectra. This "bump" is the result of the transformation from the initially higher free energy (higher vapor pressure) amorphous solid to the lower free energy (lower vapor pressure) crystalline phase.[9,16,29] The "bump" in the desorption spectrum is a signature of the crystallization of the ASW film.

The middle set of curves in Figure 1 is for the experiment with a wait time of 3600 s. These curves are similar to those for the 0 s wait time experiment except that the onset of desorption from the bottom layer ($H_2^{16}O$, red curve) begins at a lower temperature of about 150 K. The top set of curves is for the experiment with a wait time of 10800 s. In this case, the spectra show that the desorption curves for the two isotopes are nearly the same at 140 K which indicates that the two isotopic layers have completely intermixed by this temperature.

The experiments show that the intermixing of the isotopic layers increases with wait time and thus provides evidence for translation motion at 136 K. If there was no translational motion, we would expect that the sets of TPD spectra for all the experiments to have the same behavior, independent of the wait time. It is important to note that the use of nanoscale films and long wait times are requisite to observe translational motion at the extremely low diffusivities ($10^{-20}$ - $10^{-22}$ m$^2$/s) at temperatures near and just above $T_g$.

While the results in Figure 1 provide clear, albeit qualitative, evidence for translational diffusion at 136 K, extracting quantitative diffusivities from these types of "wait and flash" experiments can be complicated. One complication is that, while diffusion occurs at 136 K during the isothermal wait time, it also occurs during the heating ramp. Extracting quantitative information requires decoupling the two sources of diffusion. Another complication is that even at low temperatures, there is desorption from the top layer of the composite film. For example, at 136 K the estimated desorption rate for amorphous water is ~1.2 × $10^{-3}$ ML/s.[30] This means that there can be significant desorption for the longer wait time experiments. For example, for a wait time of 10800 s, at 136 K there should be the loss of ~13 ML from the film. This is consistent with the experimental results in Figure 1 and is another factor to consider when trying to quantify the diffusivity.

In our prior work using isotopic ASW layers, the observed intermixing occurred at temperatures (150 – 155 K) in concert with the crystallization of the film.[15,25,26] In those experiments the films were heated at 0.6 K/s. This required us to model the diffusivity as a linear combination of the amorphous and crystalline diffusion weighted by their respective mole fractions. The close coupling of diffusion and crystallization means that extracting the true translational liquid diffusion from intermixing measurements is complicated. Others have shown that additional mechanisms can arise such as mixing through and along cracks that form when ASW crystallizes.[31-34] Here, to simplify the analysis and to eliminate potential specious diffusion mechanisms we conducted experiments using very slow heating rates to decouple the diffusion and crystallization kinetics. Decoupling the diffusion and crystallization kinetics requires that the activation energies for diffusion and crystallization are different. If that is the case, it should be possible to observe intermixing at temperatures that are below, and well-separated from crystallization.

The effect of slowing the heating rate on the intermixing is displayed in Figure 2. Plotted are the results for three experiments for ASW films composed of 20 layers of $H_2^{16}O$ deposited on top of 80 layers of $H_2^{18}O$ grown at 70 K and then heated with a linear ramp rate. The displayed spectra were obtained with a heating rate of 0.01 K/s (bottom set), of 0.001 K/s (middle set), and of 0.0001 K/s (top set). As expected, the desorption curves shift to lower temperatures with decreasing heating rate. This is due to the activated kinetic processes (desorption, crystallization, and diffusion) having more time to progress at lower temperatures. Note that the use of slow heating rates comes at the cost of a decrease in signal intensity. The spectra show that there is about a ten-fold decrease in signal intensity for every factor of ten decrease in heating rate. However, even for the slowest heating rate of 0.0001 K/s the signal to noise is still very good. This is due to the use of molecular beam deposition, line of sight mass spectrometer detection, and increased signal averaging times. For all heating rates, signals were averaged to give a temperature resolution greater than 0.01 K.

The purpose of using very slow heating rates is to decouple the diffusion and crystallization kinetics. The arrows in Figure 2 mark the temperatures where the desorption rates of the two isotopes are equal which we will refer to as the "cross-over" temperature. The cross-over temperature is an indicator for the intermixing of the isotopic layers and thus the diffusivity. For all the heating rate experiments in Figure 2 the "cross-over" temperature is well-separated from the onset of crystallization ("the bump"). However, due to differences between the activation energies for diffusion and crystallization, the temperature shifts are not the same for the two processes. For example, for the 0.01 K/s experiment the temperature difference between the crossover temperature and the crystallization bump is ~3 K whereas for the 0.0001 K/s experiment the difference is ~9 K. The reason is that the slower ramp rate experiment spends more time at lower temperatures which allows for the lower activation energy diffusion process to proceed relatively faster than the higher activation energy crystallization process. The effect is similar to that observed in Figure 1.

Figure 3 shows a plot of the diffusive mixing (cross over, solid green circles) and crystallization (solid black circles) temperatures for the experiments in Figure 2 along with the data from two additional heating rate experiments (0.03 and 0.003 K/s). The plot shows that for heating rates below ~0.2 K/s diffusive intermixing occurs at a lower temperature than crystallization. For heating rates greater than ~0.2 K/s the reverse is true, that is, crystallization occurs at a lower temperature than diffusive intermixing. The difference between the diffusive intermixing and crystallization temperatures increases with decreasing heating rates. This is a result of the activation energy for diffusion being less than that for crystallization as supported by the slopes of the diffusion (green line) and crystallization (black line) curves in Figure 3. Clearly, using slow heating rates will allow us to measure the translational liquid diffusivity without interference from ASW crystallization.

Figure 4 displays the experimental results (solid lines) for the 0.0003, 0.001, and 0.003 K/s experiments (all heating rate experiments are shown in the SI). Analysis of the experimental results was conducted using numerical simulations with a one-dimensional kinetic model where the system was divided into N discrete layers corresponding to the number of layers of the ASW films. We assumed the initial layer concentrations to be 100% $H_2^{16}O$ in the top 20 ML and 100% of the $H_2^{16}O$ isotope in the bottom 80 ML. Diffusion between the adjacent layers (the layer above and below), was governed by the isotopic concentration gradient between layers and the temperature-dependent diffusivity. Molecules in the top layer were allowed to desorb with a rate dependent on the isotopic mole fraction, the desorption kinetic parameters, and the phase of the film (amorphous or crystalline). The crystallization kinetics were modeled with a using a differential form of the Avrami equation.[15,29] The total desorption rate is the sum of the amorphous and crystalline desorption rates weighted by their respective mole fractions. The diffusivity was also weighted by the crystalline mole fraction however the diffusivity for the crystalline phase was set to zero which froze in the isotope distribution when the film had completely crystallized. An Arrhenius temperature dependence, *rate = v exp(−E$_A$/RT)*, where *v* is the prefactor and $E_A$ is the activation energy, was used in the simulations for all kinetic processes. The full list of diffusivity parameters is given in the SI. Note that the diffusivities we report below were extracted from desorption data prior to crystallization.

The results for a set of simulations using the same diffusion activation energy are displayed in Figure 4 as dashed lines. The simulations are in good agreement with the experimental results and quantitatively capture the desorption, crystallization, and diffusion kinetics. An activation energy for diffusion of $E_A$ = 36.82 kJ/mol was used for all the heating rate experiments displayed in Figure 4. The diffusion prefactors varied slightly with *v* =1.5, 1.4 and 1.2 × $10^{-6}$ m$^2$/s used for the 0.003, 0.001, and the 0.0003 K/s simulations respectively in Figure 4. However, due to compensation effects equally reasonable simulation results fits can be obtained with a range of diffusion activation energy and prefactor combinations. A series of simulations were conducted using a range of diffusion activation energies. The optimal diffusion activation energy was determined by the value that gave the smallest average standard deviation for the set of prefactors needed to simulate the five heating rate experiments. The optimal diffusion activation energy was determined to be $E_A$ = 36 ± 4 kJ/mol and the corresponding diffusion prefactors to be $\nu = 10^{-6.17 \pm 1.58}$ m²/s. (See SI for more details.)

Figure 5 displays an Arrhenius plot of the temperature dependent diffusivity obtained in the current paper (solid black line) and our estimated uncertainties (dashed black lines). The vertical dashed line at ~147 K is the crystallization temperature for the 0.01 K/s experiment. Above this temperature crystallization impacts the mixing kinetics.[27,28] We have previously estimated the diffusivity in ASW films using the crystalline growth rate.[35] In those experiments an ASW layer was deposited on top of a crystalline ice layer and then heated to, and held isothermally at a temperature between 126 and 151 K. In this configuration the crystalline ice layer acts as a nucleation template and the crystallization of the ASW layer is entirely due to a linear growth front.[35-37] The crystalline growth rate of a material can be related to the liquid mobility of molecules at the liquid–solid interface using the Wilson-Frenkel model.[38-42] While the Wilson-Frenkel model does not provide a direct measurement of diffusivity, it is based on reasonable physical arguments and its application often works well.[41,43] The diffusivities obtained from the crystalline ice growth rates and the Wilson-Frenkel model are shown in Figure 5 as solid blue circles.[35] There is good agreement between the diffusivities obtained with the intermixing of isotopic layers (present work) and those extracted using the crystalline ice growth rates. Notice that above ~145 K those diffusivities begin to be higher than our diffusion estimate (solid black line). This is the temperature above which the diffusion activation energy was observed to increase, indicating a strong (Arrhenius) to fragile (super-Arrhenius) transition.[35] Our measured diffusivities are limited to the low temperature, strong temperature dependence regime below 145 K.

In previous work we used the permeation of an inert gas through an amorphous overlayer to measure the diffusivity of various supercooled liquids prior to crystallization.[44-48] In that work we developed universal scaling relationships between the diffusivity, overlayer thickness, and the temperature ramp rate. Using these scaling relationships, we derived simple equations from which the diffusivity can be extracted using the inert gas peak desorption temperature, heating rate, and layer thickness without the need for numerical simulation.[47] Those same scaling relationships and equations were employed using the "cross-over" temperature as a characteristic reference to extract the supercooled liquid diffusivity for the TPD experiments here. The results from the scaling analysis are in good agreement with the simulation results in Figure 5 and are shown in the SI (Figure SI-12).

The experimental diffusivities in this paper unambiguously demonstrate translational liquid-like motion at temperatures near and above the proposed $T_g$ of 136 K. Our results are in contrast to the interpretations of Fisher and Devlin[23] where they claim that there is no evidence for translation motion in the observed isotope exchange was only due to rotations. It is unclear if the experiments of Fisher and Devlin[23] are only sensitive to rotations and not translations as claimed. Similarly, Shepard and Salzman[24] based their conclusions on calorimetric measurements of the heat capacity increase near the glass transition temperature. While these observations are interesting and important, it is not clear that they can be used to exclude translational diffusion. Our experiments were specifically designed to probe only translational motion. It is easy to envision water molecules exhibiting rotation without translation, but it is difficult to imagine how they can translate without rotating due to the strongly anisotropic nature of hydrogen bonds in condensed phases. One possibility is that rotations are faster than translations. Resolving this will require additional research beyond the scope of this paper.

Our results also support the interpretation that the relatively small increase in the heat capacity of ~2 J/K·mol near 136 K as being due to the glass transition.[17-20] Arguments against this interpretation are based on the fact that the small change in the heat capacity reported for amorphous water is only ~2% of that observed for water solutions.[2,49] For most systems there is a relatively large abrupt increase the heat capacity at the glass transition temperature as previously "frozen" degrees of freedom (rotations, translations) become accessible. This expectation assumes that the structures of the amorphous solid and supercooled liquid are very different. However, more recent work has shown that the structure of supercooled liquid water is composed of a mixture of low- and high-temperature structural motifs.[50-52] The low-temperature motif is more "ice-like" with a high level of local tetrahedral configurations whereas the high-temperature motif is akin to the normal liquid above the melting temperature. The composition of supercooled liquid water varies smoothly with temperature with the low temperature structure being dominant below 135 K and the high-temperature structure being dominant above 245 K. This may explain why near $T_g$ the transition from the highly coordinated amorphous solid to a supercooled liquid whose structure also includes a high degree of coordination does not result in a large increase in the heat capacity.

## II. Conclusions

The experimental results on the intermixing of isotopically layered nanoscale ASW films presented here provide direct and conclusive evidence for translational liquid-like water diffusion at temperatures near 136 K. The observations required very slow heating rates to allow for sufficient time at low temperatures to observe diffusion and to decouple the diffusion and crystallization kinetics. The diffusivities were obtained over a temperature range from 125 to 145 K and are consistent with those obtained using the crystal ice growth rate and the Wilson-Frenkel model. These findings support the idea that the melt of amorphous water above its glass transition temperature is thermodynamically continuous with normal supercooled liquid water near the melting point.

## III. Experimental Methods

The experiments were performed in an ultrahigh vacuum UHV chamber (base pressure $<1.0 \times 10^{-10}$ Torr that has been described before in detail.[53] Briefly, the ASW films were created by deposition from a quasi-effusive molecular beam onto a graphene covered 1 cm diameter Pt(111) sample at normal incidence at a temperature of 70 K. The molecular beam just overfills the sample which helps to minimize the background signal and increase the detection sensitivity. The sample was spot welded to two tantalum wire leads clamped in a Cu jig attached to a closed cycle He cryostat and resistively heated through the Ta leads. The temperature was monitored with a K-type thermocouple spot welded to the back of the sample and controlled by computer. Temperature was calibrated using the amorphous water desorption rates calculated from the published vapor pressure data.[30] The error in the absolute temperature was estimated to be less than $\pm 2$ K. The TPD spectra of the deposited films were obtained with a quadrupole mass spectrometer in a line-of-sight configuration about 1 cm from the sample. The line-of-sight configuration increases the signal intensity by about a factor of ten compared to background desorption detection methods.

## Acknowledgment

This work was supported by the U.S. Department of Energy (DOE), Office of Science, Office of Basic Energy Sciences, Division of Chemical Sciences, Geosciences, and Biosciences (FWP 16248).

**Supporting Information**

Supporting Information is available free of charge and contains plots of the temperature programmed desorption (TPD) and simulation (for $E_A$ = 36.82 kJ/mol) results for all five of heating rate experiments displayed on linear-linear and semi-log plots (S1-S10). Simulation diffusion prefactors (ν) and activation energies ($E_A$) that accurately predict the "crossover temperature" for the mixing of the isotopic layers are presented in Table SI-1. The analysis of these parameters to estimate the optimal set of diffusion parameters is shown in Figure SI-11. Figure SI-12 shows a comparison of the supercooled liquid water diffusivity obtained from numerical simulations described in the main manuscript with those obtained using equations derived from scaling relationships.

**Figure Captions**

**Figure 1** Experimental TPD spectra for amorphous water films composed of 20 layers of $H_2^{18}O$ (blue curve) deposited on top of 20 layers of $H_2^{16}O$ (red curve) at 70 K. After deposition, the composite films were heated to and held at 136 K for various wait times 0 s (bottom), 3600 s (middle), and 10800 s (top). After the specific wait time, the films were cooled below 70 K and then heated at a linear rate of 0.5 K/s until desorbed.

**Figure 2** Experimental TPD spectra for amorphous water films composed of 20 layers of $H_2^{16}O$ (red line) deposited on top of 80 layers of $H_2^{18}O$ (blue line) at 20 K. After deposition, the composite films were heated at rates of 0.01 K/s (bottom set), 0.001 K/s (middle set) and 0.0001 K/s (top set) until completely desorbed. The horizontal arrows mark the temperature where the desorption rates of the two water isotopes are equal ("cross-over" temperature) which is an indication that the films have intermixed.

**Figure 3** Plot of the diffusive mixing (solid green circles) and crystallization (solid black circles) temperatures versus heating rate for the slow heating rate experiments using amorphous water films composed of 20 layers of $H_2^{16}O$ deposited on top of 80 layers of $H_2^{18}O$. The heating rates used were 0.01, 0.003, 0.001, 0.0003, and 0.0001 K/s. The diffusive mixing (crossover) temperature is defined as the temperature where the desorption rates of the two isotopes are equal. The crystallization temperature is defined as when the film is 50% crystallized which is determined by analysis of the "bump" region of the TPD spectrum.

**Figure 4** Experimental TPD spectra for amorphous water films composed of 20 layers of $H_2^{16}O$ (solid red lines) deposited on top of 80 layers of $H_2^{18}O$ (solid blue lines) at 20 K. After deposition, the composite films were heated at rates of 0.003 K/s (bottom set), 0.001 K/s (middle set), and 0.0003 K/s (top set) until completely desorbed. The dashed lines are the corresponding simulation results obtained with diffusion Arrhenius parameters of $E_A$ = 36.82 kJ/mole and prefactors of $v$ =1.5, 1.4 and 1.2 × $10^{-6}$ $m^2$/s for the 0.003, 0.001, and the 0.0003 K/s simulations.

**Figure 5** Arrhenius plot of the estimated temperature dependent diffusivity obtained in the current paper (solid black line) ($E_A$ = 35.98 kJ/mol and $v$ = 6.6 × $10^{-7}$ $m^2$/s) and our estimated uncertainties (dashed black lines). The diffusivities obtained using the measured crystalline ice growth rates and the Wilson Frenkel model (solid blue circles) in reference 35. The red vertical dashed line at ~147 K is the crystallization temperature for the 0.01 K/s experiment which demarks the highest temperature where our diffusivity measurements were directly measured.

Figure 1

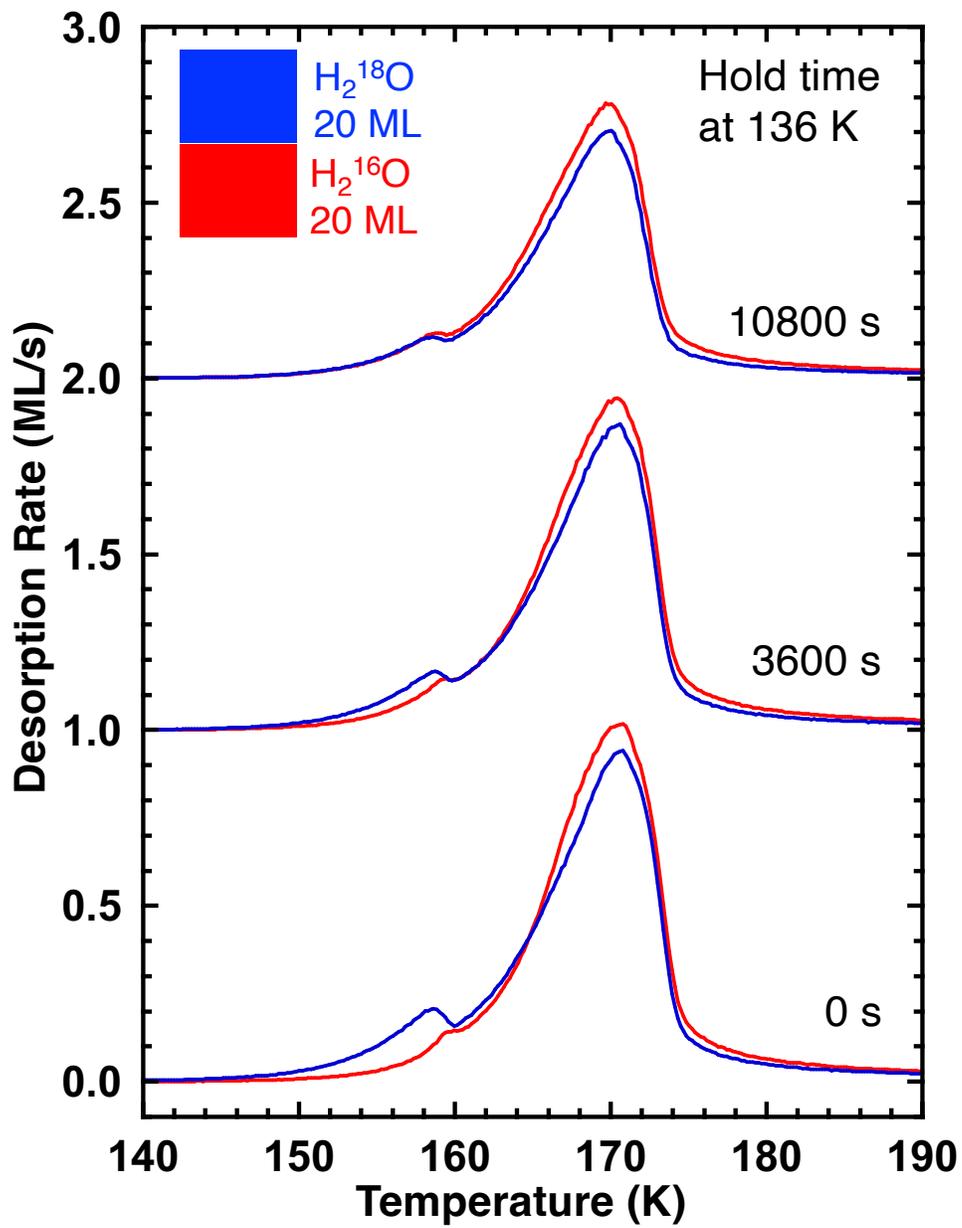

Figure 2

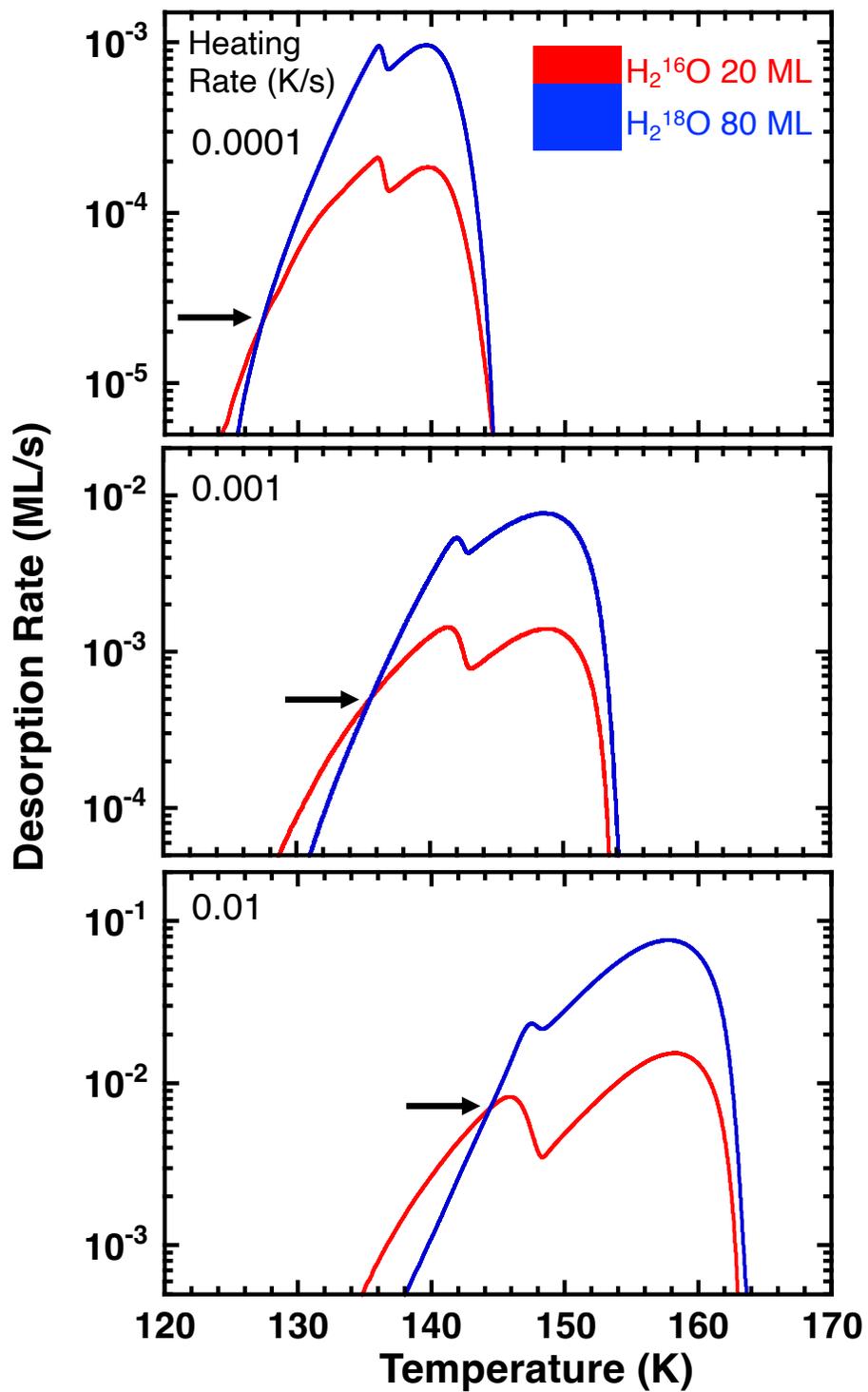

Figure 3

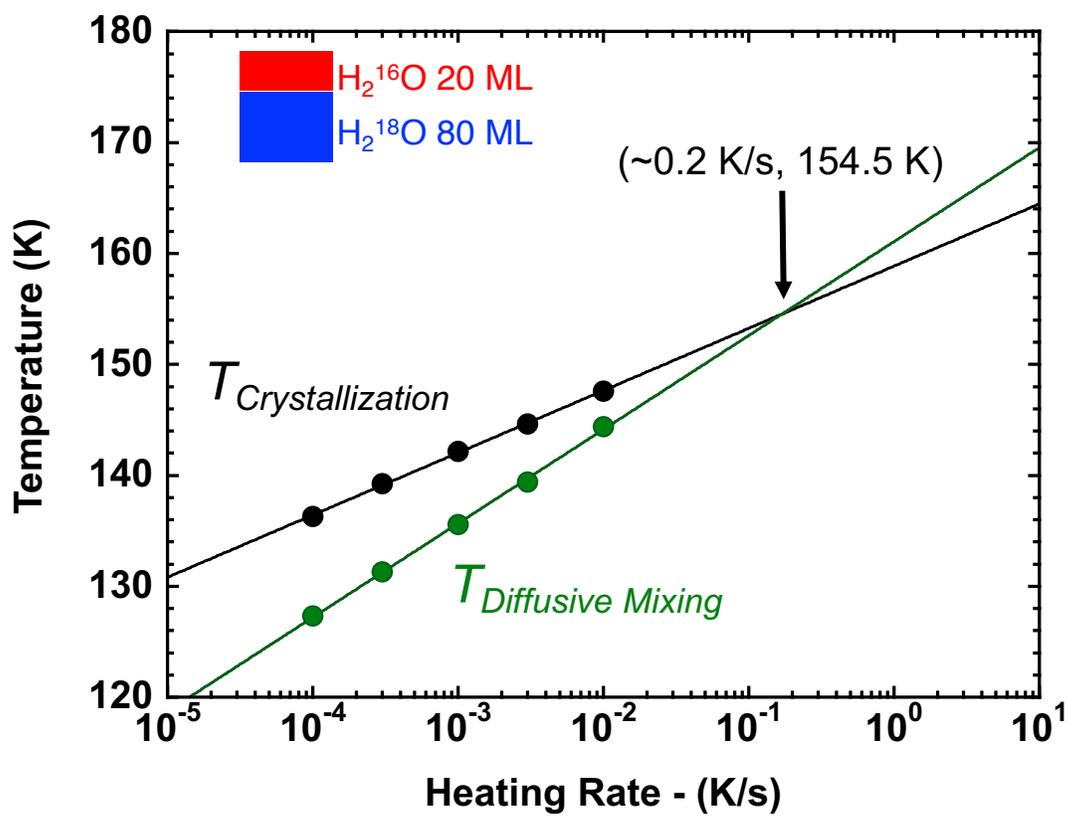

Figure 4

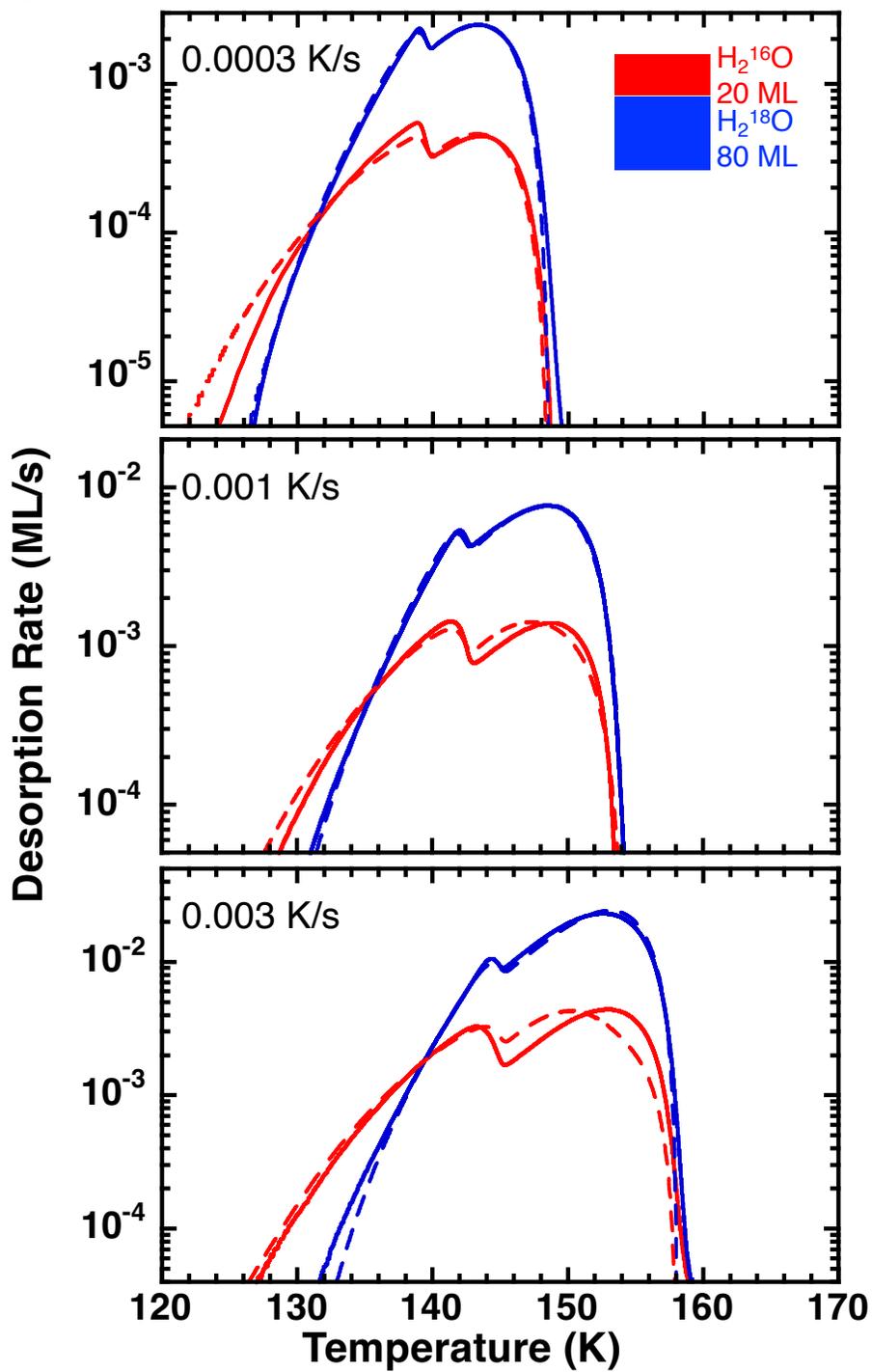

Figure 5

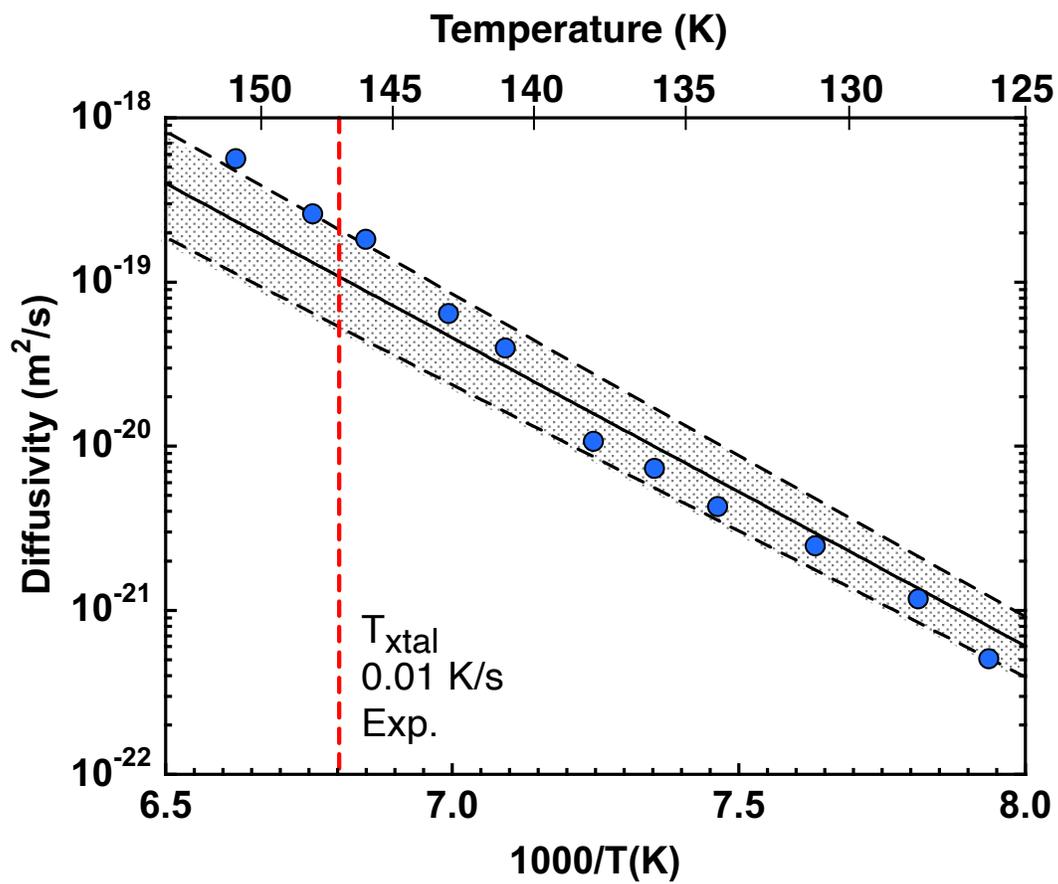

**Supporting Information for:**

**Supercooled Liquid Water Diffusivity at Temperatures near the Glass Transition Temperature**

R. Scott Smith, Wyatt Thornley, Greg A. Kimmel, and Bruce D. Kay

*Physical Sciences Division, Pacific Northwest National Laboratory, Richland, Washington 99352*

R. Scott Smith, Email: Scott.Smith@pnnl.gov

Bruce D. Kay, Email: Bruce.Kay@pnnl.gov

This Supporting Information document contains plots of the temperature programmed desorption (TPD) results for all five of heating rate experiments displayed on linear-linear and semi-log plots (S1-S10). The heating rates were 0.0001, 0.0003, 0.001, 0.003, and 0.01 K/s. The simulation results using a diffusion activation energy of 36.82 kJ/mol are also displayed in these plots. These figures support the points made in the main text but for reasons of brevity and clarity were not included there.

Simulation diffusion prefactors ($\nu$) and activation energies ($E_A$) that accurately predict the "crossover temperature" for the mixing of the isotopic layers are presented in Table SI-1). For a given $E_A$, slightly different prefactors were needed to fit each of the five heating rate experiments. We used the percent standard deviation in the prefactors for a given activation energy to estimate the "best" set of diffusion parameters. Figure SI-11 shows a plot of this analysis.

Figure SI-12 shows a comparison of the supercooled liquid water diffusivity obtained from numerical simulations described in the main manuscript with those obtained using equations derived from scaling relationships we have previously published.[1] The data used for the scaling equation calculations is contained in Table SI-2.

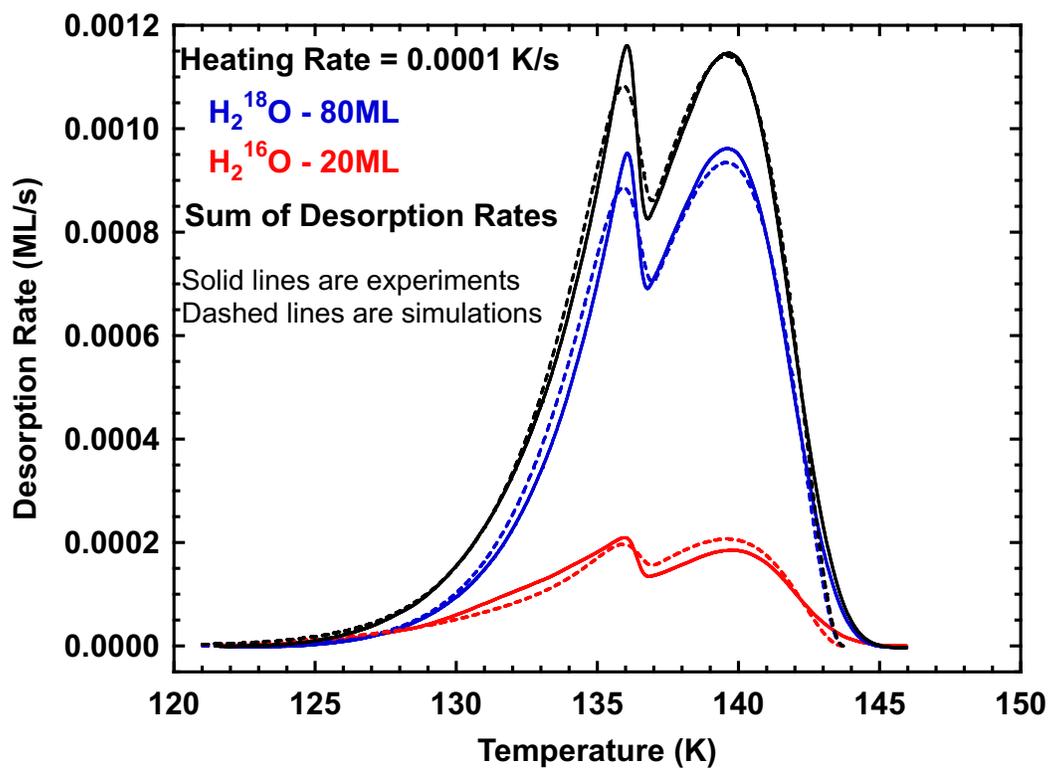

**Figure SI-1.** Experimental TPD spectra for amorphous water films composed of 20 layers of $H_2^{16}O$ (solid red lines) deposited on top of 80 layers of $H_2^{18}O$ (solid blue lines) at 20 K. After deposition, the composite film was heated at a rate of 0.0001 K/s. The dashed lines are the corresponding simulation results obtained with diffusion Arrhenius parameters of $E_A$ = 36.82 kJ/mol and prefactor of $1.60 \times 10^{-6}$ m$^2$/s. The solid black line is the sum of the $H_2^{16}O$ and $H_2^{18}O$ desorption rates and the solid line is the simulation result. The agreement between the total desorption rate and the simulation confirms the accuracy of the desorption and crystallization kinetic parameters.

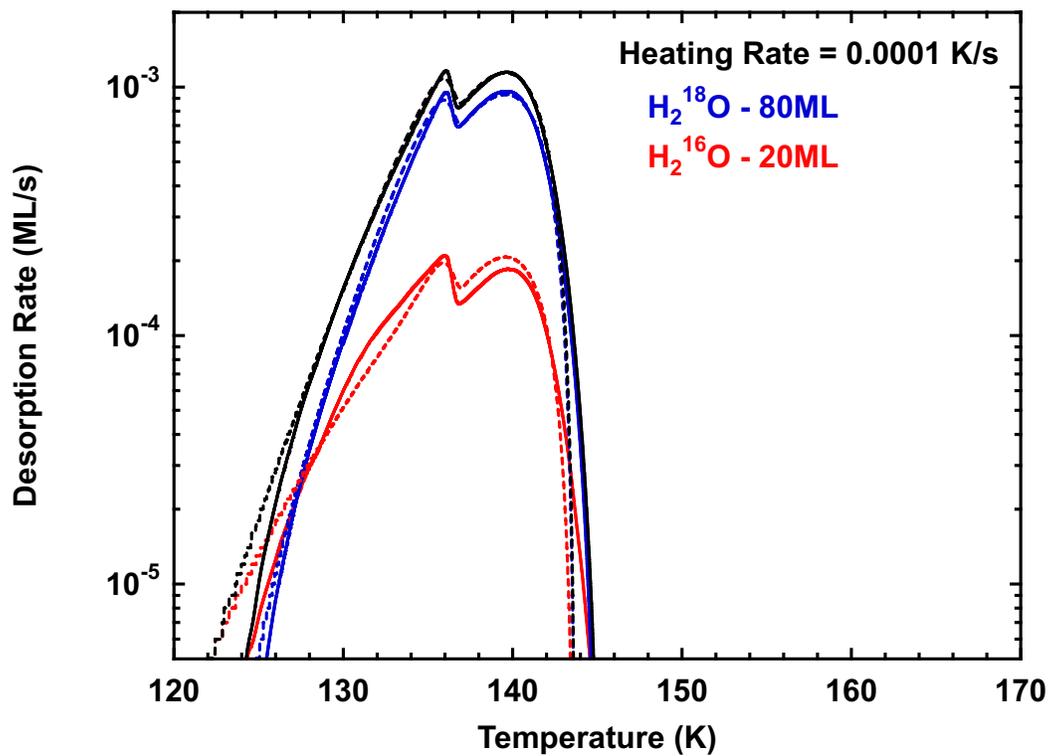

**Figure SI-2.** The desorption (solid lines) and simulation (dashed line) results for the 0.0001 K/s heating rate experiment from Figure SI-1 displayed on a semi-log plot.

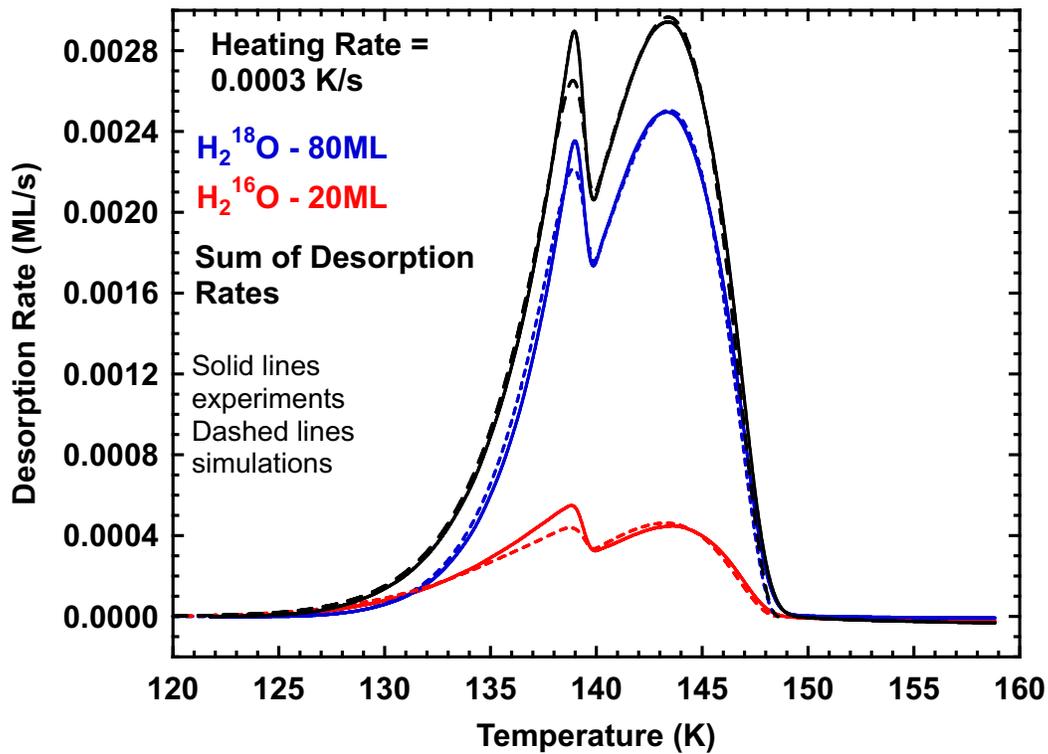

**Figure SI-3.** Experimental TPD spectra for amorphous water films composed of 20 layers of $H_2^{16}O$ (solid red lines) deposited on top of 80 layers of $H_2^{18}O$ (solid blue lines) at 20 K. After deposition, the composite film was heated at a rate of 0.0003 K/s. The dashed lines are the corresponding simulation results obtained with diffusion Arrhenius parameters of $E_A$ = 36.82 kJ/mol and prefactor of $1.20 \times 10^{-6}$ m$^2$/s. The solid black line is the sum of the $H_2^{16}O$ and $H_2^{18}O$ desorption rates and the solid line is the simulation result. The agreement between the total desorption rate and the simulation confirms the accuracy of the desorption and crystallization kinetic parameters.

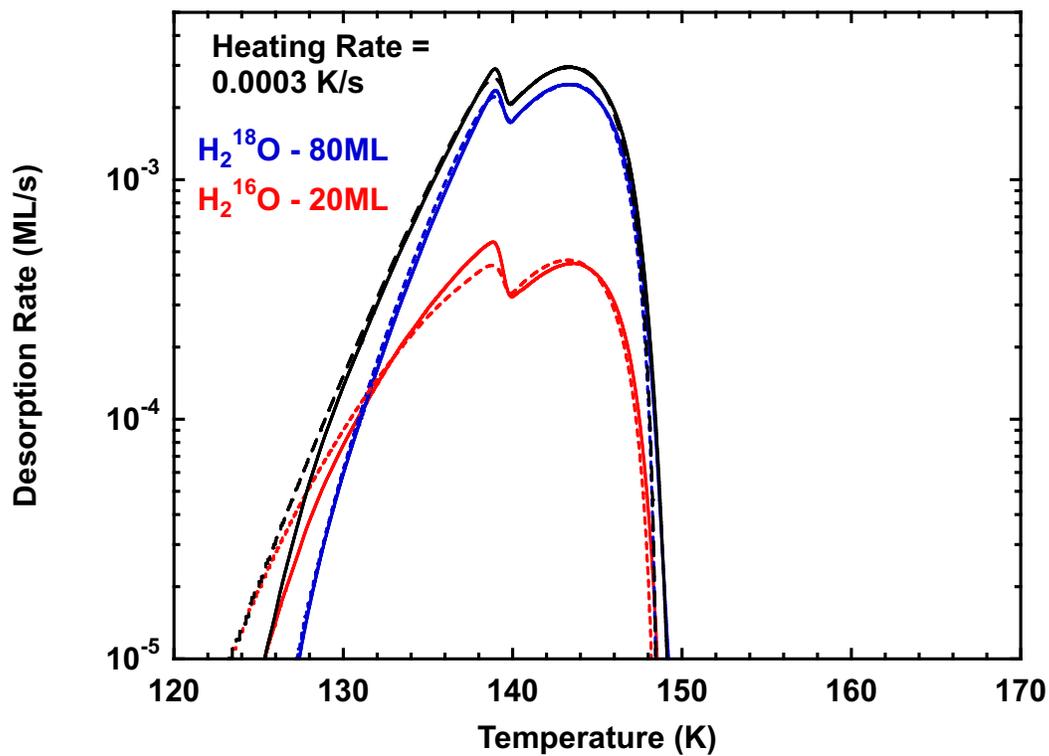

**Figure SI-4.** The desorption (solid lines) and simulation (dashed line) results for the 0.0003 K/s heating rate experiment from Figure SI-3 displayed on a semi-log plot.

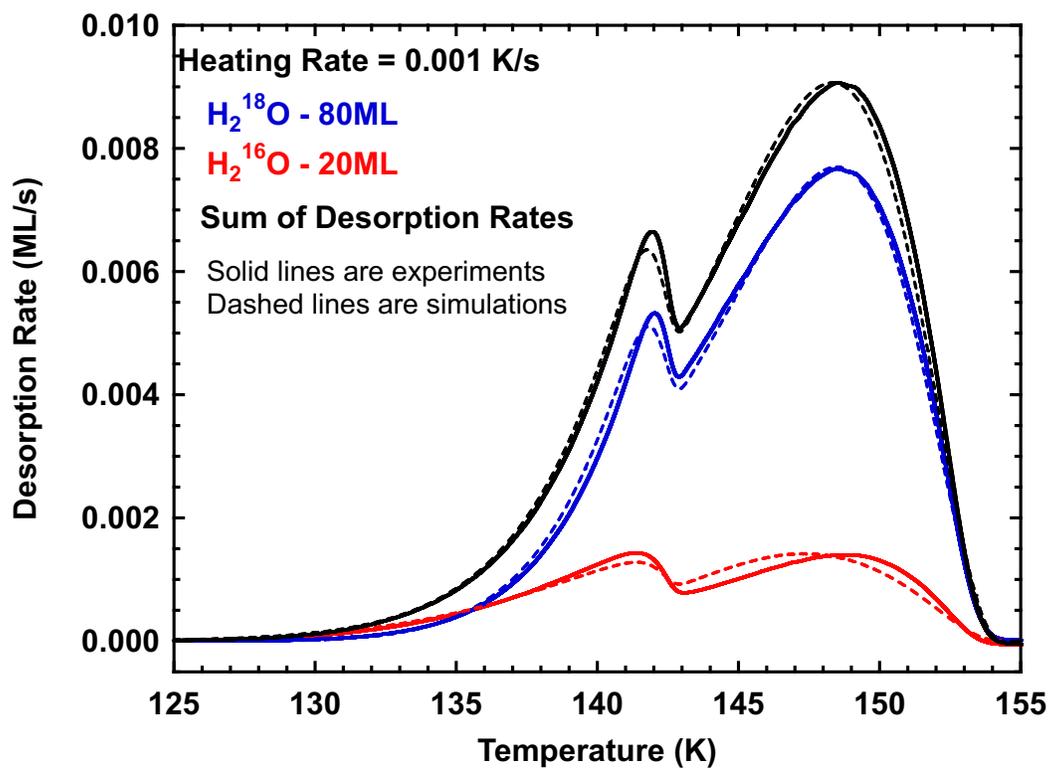

**Figure SI-5.** Experimental TPD spectra for amorphous water films composed of 20 layers of $H_2^{16}O$ (solid red lines) deposited on top of 80 layers of $H_2^{18}O$ (solid blue lines) at 20 K. After deposition, the composite film was heated at a rate of 0.001 K/s. The dashed lines are the corresponding simulation results obtained with diffusion Arrhenius parameters of $E_A$ = 36.82 kJ/mol and prefactor of $1.40 \times 10^{-6}$ m$^2$/s. The solid black line is the sum of the $H_2^{16}O$ and $H_2^{18}O$ desorption rates and the solid line is the simulation result. The agreement between the total desorption rate and the simulation confirms the accuracy of the desorption and crystallization kinetic parameters.

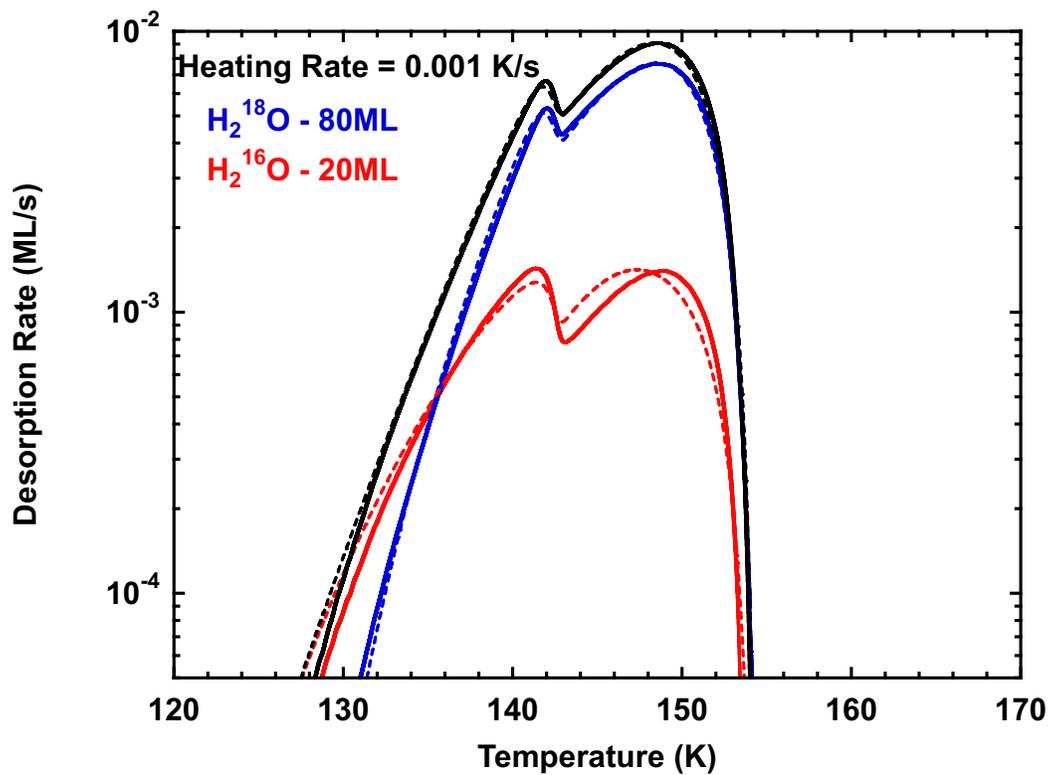

**Figure SI-6.** The desorption (solid lines) and simulation (dashed line) results for the 0.001 K/s heating rate experiment from Figure SI-5 displayed on a semi-log plot.

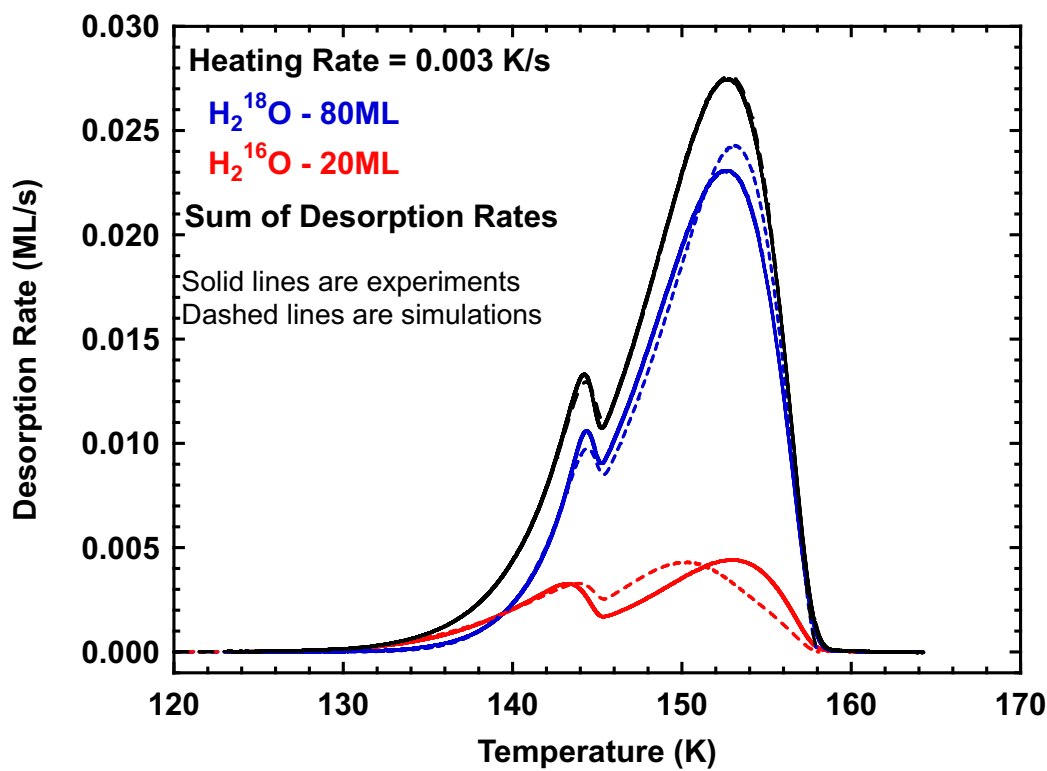

**Figure SI-7.** Experimental TPD spectra for amorphous water films composed of 20 layers of $H_2^{16}O$ (solid red lines) deposited on top of 80 layers of $H_2^{18}O$ (solid blue lines) at 20 K. After deposition, the composite film was heated at a rate of 0.003 K/s. The dashed lines are the corresponding simulation results obtained with diffusion Arrhenius parameters of $E_A$ = 36.82 kJ/mol and prefactor of $1.50 \times 10^{-6}$ m²/s. The solid black line is the sum of the $H_2^{16}O$ and $H_2^{18}O$ desorption rates and the solid line is the simulation result. The agreement between the total desorption rate and the simulation confirms the accuracy of the desorption and crystallization kinetic parameters. At this higher heating rate, the agreement between experiment and simulation begins to break down above the crystallization temperature. This is likely due to other diffusion/intermixing processes that occur during ASW crystallization that are not the result of true translational liquid diffusion.

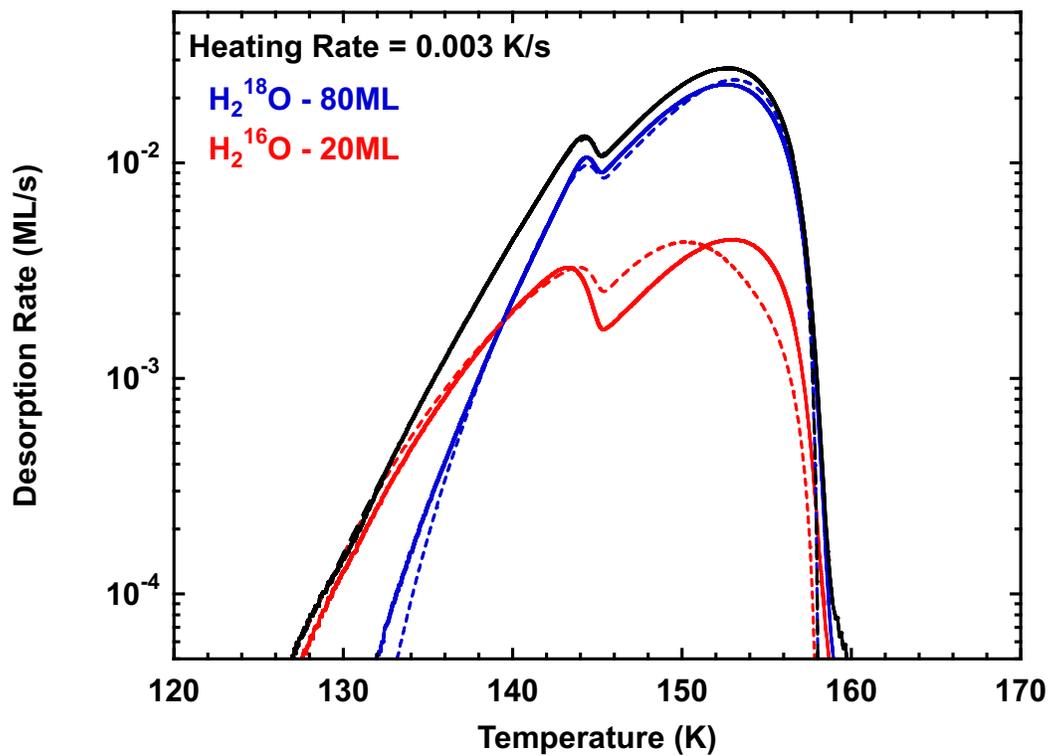

**Figure SI-8.** The desorption (solid lines) and simulation (dashed line) results for the 0.003 K/s heating rate experiment from Figure SI-7 displayed on a semi-log plot.

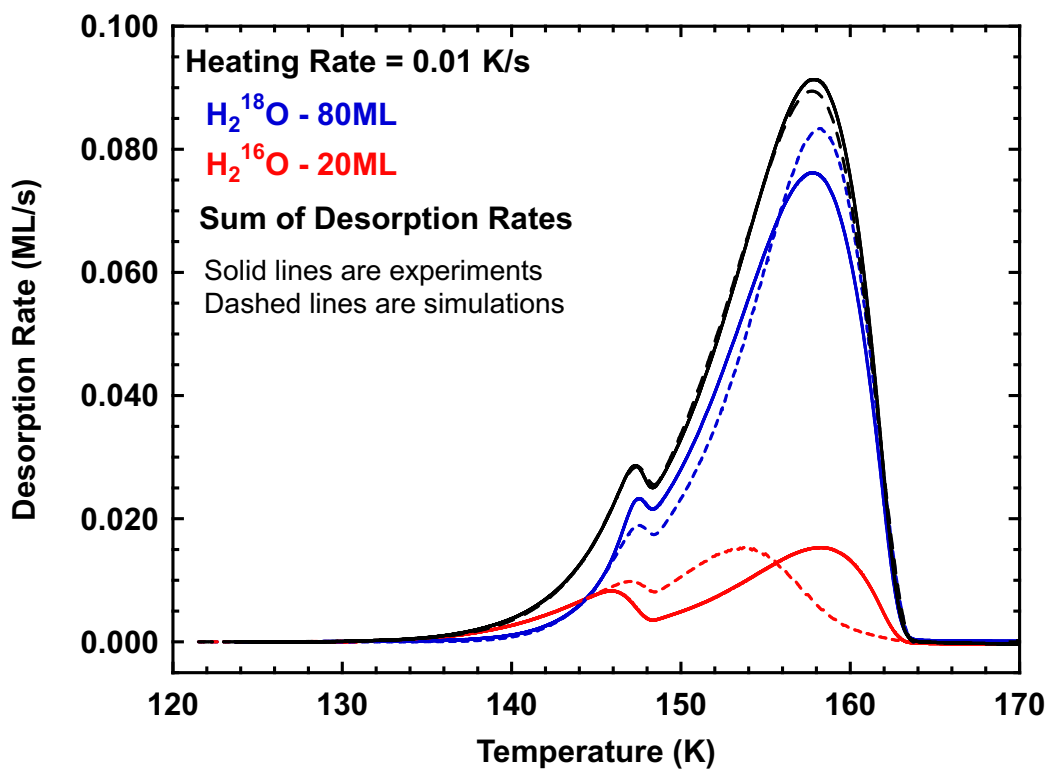

**Figure SI-9.** Experimental TPD spectra for amorphous water films composed of 20 layers of $H_2^{16}O$ (solid red lines) deposited on top of 80 layers of $H_2^{18}O$ (solid blue lines) at 20 K. After deposition, the composite film was heated at a rate of 0.01 K/s. The dashed lines are the corresponding simulation results obtained with diffusion Arrhenius parameters of $E_A = 36.82$ kJ/mol and prefactor of $1.35 \times 10^{-6}$ m²/s. The solid black line is the sum of the $H_2^{16}O$ and $H_2^{18}O$ desorption rates and the solid line is the simulation result. The agreement between the total desorption rate and the simulation confirms the accuracy of the desorption and crystallization kinetic parameters. At this higher heating rate the agreement between experiment and simulation begins to break down above the crystallization temperature. This is likely due to other diffusion/processes that occur during ASW crystallization that are not the result of true translational liquid diffusion.

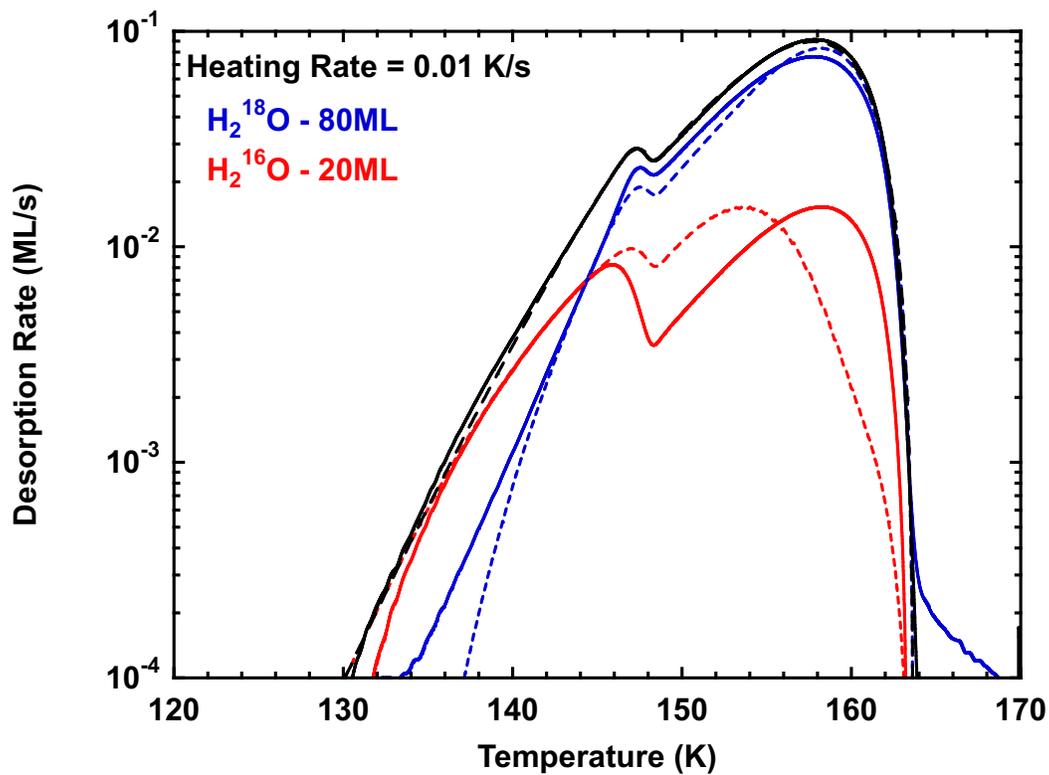

**Figure SI-10.** The desorption (solid lines) and simulation (dashed line) results for the 0.01 K/s heating rate experiment from Figure SI-9 displayed on a semi-log plot.

**Table of Diffusion Simulation Parameters**

| $E_A$ kJ/mol | ν (m²/s) 0.0001 K/s | ν (m²/s) 0.0003 K/s | ν (m²/s) 0.001 K/s | ν (m²/s) 0.003 K/s | ν (m²/s) 0.01 K/s | AVG ν (m²/s) | % Std Deviation |
|---|---|---|---|---|---|---|---|
| 29.29 | 1.15e-09 | 1.00e-09 | 1.40e-09 | 1.90e-09 | 2.10e-09 | 1.51e-09 | 31.4 |
| 33.47 | 7.70e-08 | 5.30e-08 | 6.50e-08 | 7.70e-08 | 7.70e-08 | 6.74e-08 | 14.9 |
| 36.82 | 1.60e-06 | 1.20e-06 | 1.40e-06 | 1.50e-06 | 1.35e-06 | 1.52e-06 | 10.8 |
| 39.33 | 1.90e-05 | 1.30e-05 | 1.40e-05 | 1.40e-05 | 1.20e-05 | 1.44e-05 | 18.8 |
| 41.84 | 2.00e-04 | 1.35e-04 | 1.30e-04 | 1.30e-04 | 0.95e-04 | 1.38e-04 | 27.7 |
| 34.35 | 23.00e-04 | 14.0e-04 | 12.50e-04 | 11.5e-04 | 8.5e-04 | 13.9e-04 | 39.4 |
| 46.86 | 260e-04 | 145e-04 | 125e-04 | 110e-04 | 65e-04 | 141.0e-04 | 51.6 |

**Table SI 1** Table of simulation diffusion Arrhenius parameter combinations (prefactors (ν) and activation energies ($E_A$)) that accurately predict the "crossover temperature" for the mixing of the isotopic layers. For a given $E_A$, different prefactors were needed to fit each of the five heating rate experiments. The last two columns are the average of the prefactors and the percent standard deviation for a given activation energy. While due to compensation effects, a range of activation energy/prefactor combinations can accurately predict the "crossover temperature", for a given activation energy there should be a single prefactor for all the heating rate experiments. The table shows that for a given activation energy there is a variation in the prefactor for each of the heating rate experiments. We use the magnitude of this variation to estimate the best set of diffusion Arrhenius parameters. (See next figure **SI-11).**

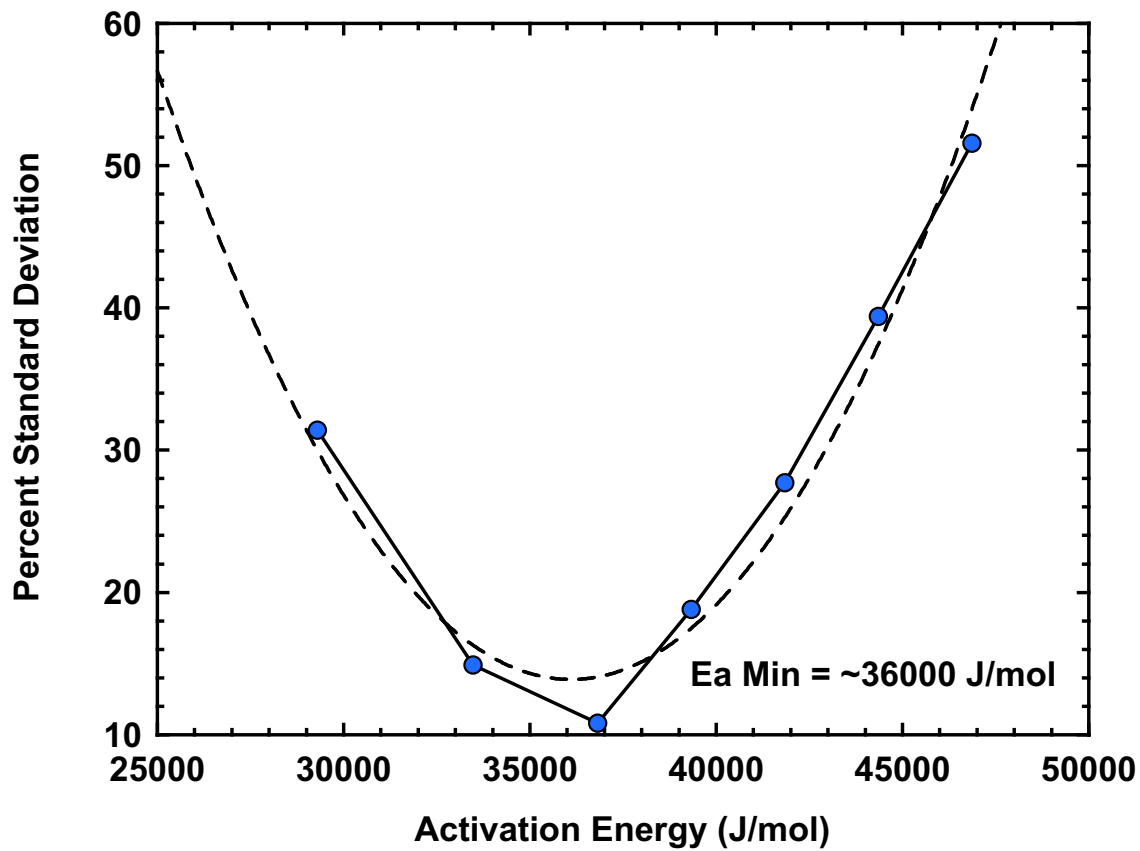

**Figure SI-11.** Plot of the percent standard deviation values of the prefactors listed in Table SI-1 (blue solid circles). The percent standard deviation is calculated relative to the average prefactor for a given activation energy. The dashed black line is a quadratic fit to the data. We estimate the diffusion activation energy to be $E_A = 36000 \pm 4000$ J/mol and the corresponding diffusion prefactors to be $v = 10^{-6.17 \pm 1.58}$ m$^2$/s.

**Comparison of numerical simulation and scaling relationship diffusivities**

In our prior work supercooled liquid diffusivities were obtained using inert gas permeation through an amorphous overlayer.[1] Scaling relationships between the diffusivity, overlayer thickness, and the temperature ramp rate were validated using kinetic simulations. From these scaling relationships we derived simple equations from which the diffusivity can be obtained using the inert gas desorption peak temperature without the need for numerical simulations. The diffusivity equation (Eq. 14 in reference 1) for those TPD experiments is $D(T_{Peak}) = \frac{\alpha \beta L^2}{T_{Peak}^2}\left(\frac{E}{R}\right)$ where $\alpha$ is a dimensionless scale factor determined by the characteristic diffusion feature being used, $\beta$ is the heating rate, $L$ is the diffusion length, $T_{peak}$ is the peak temperature in the inert gas desorption spectrum, $E$ is activation energy, and $R$ is the ideal gas constant. The ($E/R$) is determined from an Arrhenius plot of the $\beta L^2/T_{peak}^2$ term. In the present manuscript the cross over temperature, $T_x$, was used as the characteristic diffusion feature and the value for $\alpha$ was determined to be 0.30 from numerical simulations of model systems. Figure SI-12 is a plot comparing the diffusivity obtained in this paper using kinetic simulations (solid black line, see Figure 5 in main manuscript) with those obtained from scaling relationships (symbols). The plot shows that the $L = 40$ML scaling calculations (solid red circles) are in close agreement with the numerical simulation for the three slowest heating rate experiments (0.0001, 0.0003, 0.001 K/s) but begin to deviate for the faster heating rates (0.003, 0.01 K/s). The original scaling relationships were developed for experiments where the inert gas desorption occurred prior to significant desorption of the supercooled liquid overlayer. That is not the case for the present isotopic layer experiments. We can approximately account for desorption by "correcting" the value of $L$ by subtracting the amount of desorption that has occurred by $T_x$. Upon doing this, the scaling calculations (green circles) are in much better agreement with the numerical simulation. Of course, the numerical simulations take the desorption into account, but the "desorption corrected" scaling calculations are a quick and easy way to obtain a good estimate of the diffusivity without having to do a numerical simulation.

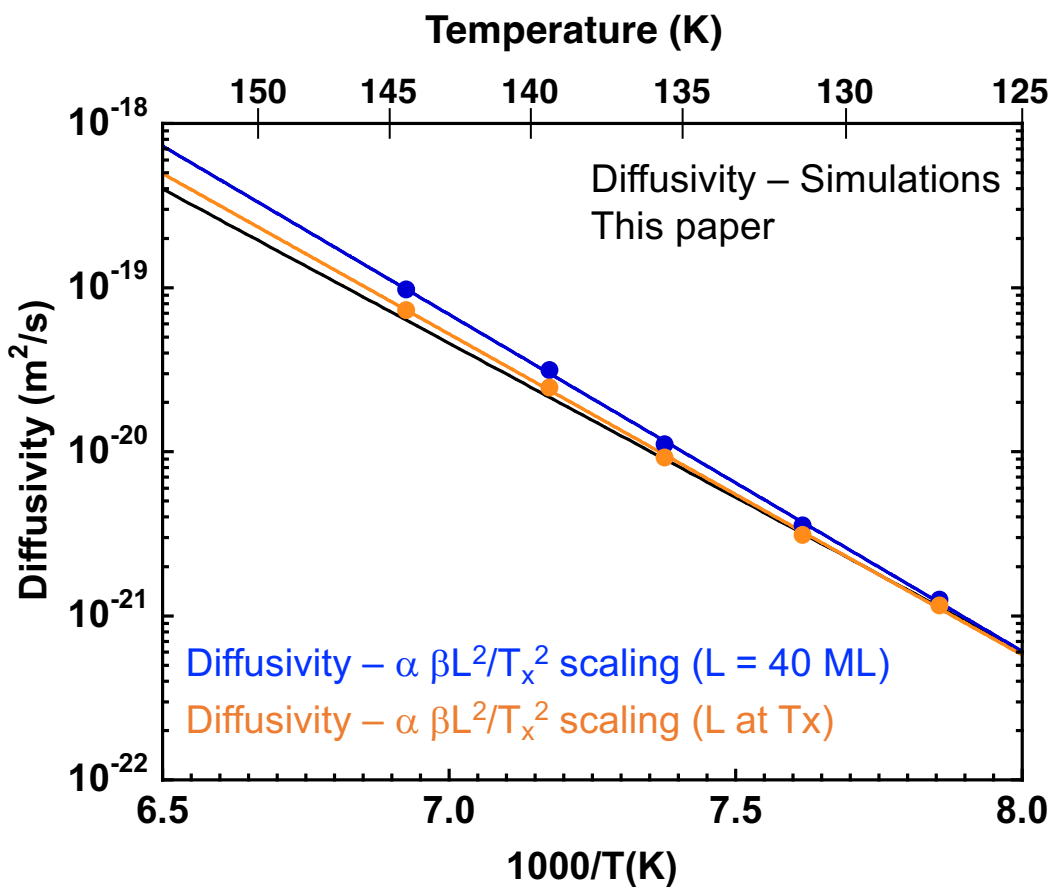

**Figure SI-12.** A plot comparing the diffusivity obtained in this paper using kinetic simulations with those obtained from scaling relationships. The solid black line is the diffusivity obtained using numerical simulations of the TPD experiments in this manuscript (see Figure 5). The diffusivities calculated using the scaling relationship $D(T_x) \propto \alpha\beta L^2/T_x^2$ are shown as solid circles. The solid blue circles are calculated using a value for $L$ of 40 ML (2 × 20 ML). The solid orange circles are calculated using a value for $L$ that is corrected for the amount desorption that has occurred at $T_x$. That is $L = (40 -$ amount desorbed at $T_x)$ ML. The simulation diffusion (solid black line) parameters are $E_A = 35.98$ kJ/mol and $v = 6.6 \times 10^{-7}$ m$^2$/s), the scaling diffusion calculated with $L = 40$ (solid blue line) parameters are $E_A = 39.30$ kJ/mol and $v = 1.6 \times 10^{-5}$ m$^2$/s), and the scaling diffusion with $L$ corrected for desorption (solid orange line) parameters are $E_A = 37.44$ kJ/mol and $v = 2.6 \times 10^{-6}$ m$^2$/s).

**Table of Parameters for Scaling Results**

| Heating rate (K/s) | $T_x$ (K) Experiment | ML Desorbed at $T_x$ | ML used for scaling equation | $T_{Crystal}$(K) Experiment |
|---|---|---|---|---|
| 0.01 | 144.40 | 4.55 | 35.45 | 147.61 |
| 0.003 | 139.38 | 3.69 | 36.31 | 144.63 |
| 0.001 | 135.58 | 2.64 | 37.36 | 142.13 |
| 0.0003 | 131.30 | 1.65 | 38.35 | 139.26 |
| 0.0001 | 127.30 | 0.75 | 39.25 | 136.28 |

**Table SI 2** Table of parameters used for the diffusivities calculated in Figure SI-12. For the calculations 1 ML =1ML = $3.0 \times 10^{-10}$ m. The table also includes both the $T_x$ and $T_{Crystal}$ temperatures given in Figure 3 in the main manuscript.